\documentclass[draft]{aipproc}
\layoutstyle{8x11single}

\usepackage{amsmath}
\usepackage{amssymb}
\usepackage{latexsym}
\usepackage{graphicx}
\usepackage{hyperref}
\usepackage{mathrsfs}
\usepackage{bm}

\newcommand{\be}{\begin{equation}}
\newcommand{\ee}{\end{equation}}
\newcommand{\ba}{\begin{eqnarray}}
\newcommand{\ea}{\end{eqnarray}}
\renewcommand{\d}{\mathrm{d}}
\newcommand{\<}{\langle}
\renewcommand{\>}{\rangle}
\newcommand{\CD}{{\cal D}}

\newcommand{\aperp}{a_{\perp }}

\newcommand{\apar}{a_{\parallel}}
\newcommand{\kapr}{\kappa(r)r^2}
\newcommand{\Hperp}{H_{\perp}}
\newcommand{\Hperpo}{H_{\perp_0}}
\newcommand{\Hpar}{H_{\parallel}}
\newcommand{\omegmo}{\Omega_{{m}}}
\newcommand{\omegko}{\Omega_{{k}}}

\newcommand{\average}[1]{\left\langle #1 \right\rangle_{\CD}}

\begin{document}

\title{On the determination of dark energy}

\author{ Chris Clarkson}{address={
Cosmology \& Gravity Group, Department of Mathematics
and Applied Mathematics, University of Cape Town, Rondebosch 7701,
South Africa.},email={chris.clarkson@uct.ac.za}}

\begin{abstract}

I consider some of the issues we face in trying to understand dark energy.  Huge fluctuations in the unknown dark energy equation of state can be hidden in distance data, so I argue that model-independent tests which signal if the cosmological constant is wrong are valuable. These can be constructed to remove degeneracies with the cosmological parameters. Gravitational effects can play an important role. Even small inhomogeneity clouds our ability to say something definite about dark energy. I discuss how the averaging problem confuses our potential understanding of dark energy by considering the backreaction from density perturbations to second-order in the concordance model: this effect leads to at least a 10\% increase in the dynamical value of the deceleration parameter, and could be significantly higher. Large Hubble-scale inhomogeneity has not been investigated in detail, and could conceivably be the cause of apparent cosmic acceleration. I discuss void models which defy the Copernican principle in our Hubble patch, and describe how we can potentially rule out these models. 

This article is a summary of two talks given at the Invisible Universe Conference, Paris, 2009.

\end{abstract}

\maketitle

\section{Introduction}

The standard LCDM model of cosmology has reached a maturity where we may think of it as a paradigm: more than just a model, it is rather a worldview underlying the  theories and methodology of our particular scientific subject. 
In many respects it is fantastic, with a handful of constant parameters describing the main features of the model, which lies neatly within the 1- or 2-$\sigma$ errors bars of most data sets. It is tempting to think of it as a science which is nearing completion, with remaining work simply reducing errors on the parameters.  However, it requires (at least) 3 pieces of physics we don't yet understand: inflation of some sort for the initial conditions; dark matter, and dark energy. 

Depending on who you speak to, the dark energy problem ranges from being not a problem at all, other than the old cosmological constant problem (landscape lovers), to being the greatest mystery/calamity in all of science~-- ever! (if you're speaking to someone writing a grant proposal). Whether it turns out to be a storm in a teacup, or a revolution in our understanding of the cosmos, it's going to be difficult getting a handle on it, and ruling out many of the possibilities may be very hard or even impossible for the foreseeable future. In this article I discuss three different aspects of the dark energy problem, and I describe some subtle observable properties of the FLRW models which can help us see if we're on the wrong track. I consider first dark energy in the exact background FLRW models, then in perturbed FLRW models, and finally discuss non-FLRW models with no dark energy in them at all.
\begin{itemize}
\item Attempting to reconstruct the dark energy dynamics poses many difficulties which are well known, not least of which are degeneracies with the other parameters of the model. The physical motivation for many models is dubious at best, yet it is clear that we have to keep an open mind about what the dynamics could be. However, there are consistency relations we may use to test specifically for deviations from $\Lambda$, which may be used without specifying a model at all. 

\item How does structure in the universe get in the way of our interpretation of a `background' model at all? That is, how do we smooth the matter in the universe to give an FLRW model? This interferes which what we think our background model should be, and hence contaminates our understanding of the dark energy. This effect is surprisingly large~-- at least 10\% in the deceleration parameter.

\item It's conceivable that dark energy has nothing to do with new physics at all, and that we're using the wrong background solutions in the first place because the universe has a large Hubble-scale inhomogeneity. These models aren't fully developed, but deserve further investigation, even though they nominally break with the Copernican principle. Again, there are observational consistency relations which let us test for and potentially falsify these sorts of deviations from the standard paradigm. 

\end{itemize}

\section{Observing $\Lambda$ -- or not}

Within the FLRW paradigm, all possibilities for dark energy can be characterised, as far as the background dynamics are concerned, by the dark energy equation of state $w(z)$~\cite{de}. Unfortunately, from a theoretical perspective the array of possibilities is vast and not well understood at all, implying that $w(z)$ could really be pretty much anything. Even Lema\^\i tre-Tolman-Bondi (LTB) void models can be interpreted, to an extent, as an effective dark energy model (see below)~\cite{ltb}. Our priority in cosmology today must therefore lie in searching for evidence for $w(z)\neq-1$. The observational challenge lies in trying to find a straightforward, yet meaningful and sufficiently general way to treat $w(z)$. Observations which explore the dynamics of the background allow us, in principle, access to two functions: $H(z)$ and $d_L(z)$, and this gives us two independent ways to establish $w(z)$. On top of this, if we are to treat dark energy as a complete unknown, the sound speed also needs to be reconstructed independently.

\begin{figure}[thbp]
\includegraphics[width=\textwidth]{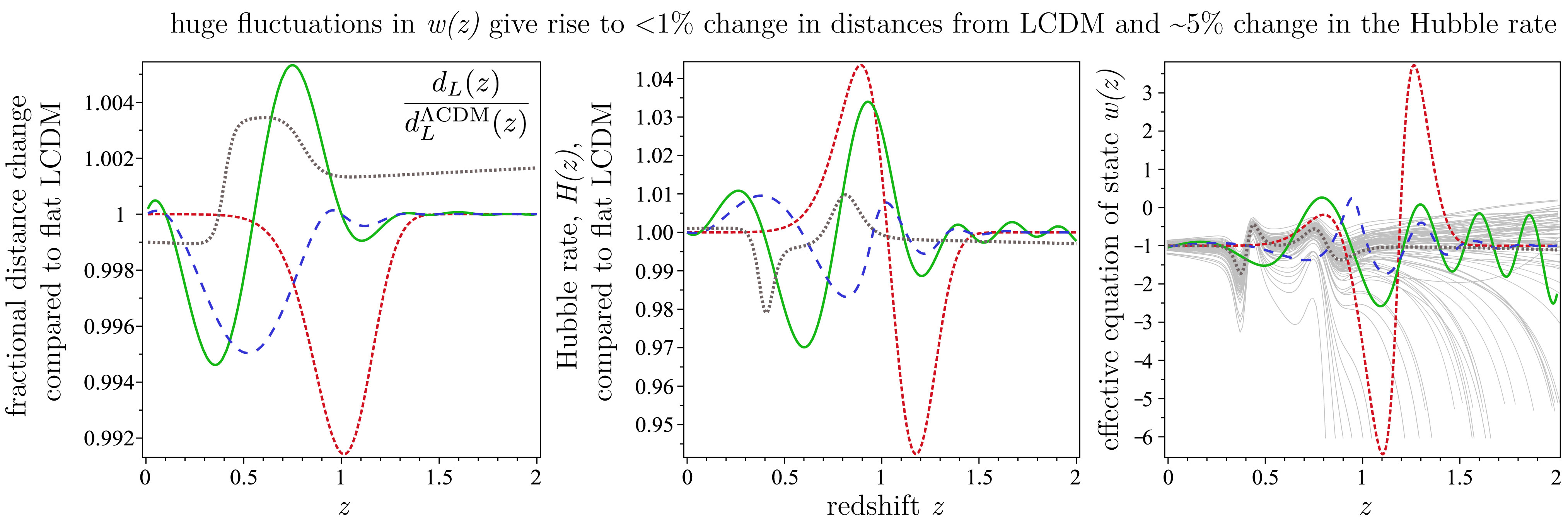}
\caption{Tiny fluctuations in $D(z)$ are amplified drastically when reconstructing $w(z)$. The fluctuations show up more strongly in $H(z)$ which is more sensitive to $w$ as an observable. This is assuming that $\Omega_m$ and $\Omega_k$ are known perfectly. In a sense then, these types of $w(z)$ are hidden in the distance data. The thin grey curves for $w(z)$ represent different reconstructions of the grey dotted curve using errors in $\Omega_m=0.3\pm0.015$ and $\Omega_k=0.0\pm0.05$~-- note that $\Lambda$ is practically consistent with this. }
\label{fig:w}
\end{figure}

The dark energy equation of state is typically reconstructed using distance measurements as a function of redshift. 
The luminosity distance may be written as 
\begin{equation}\label{d_L}
d_{L}(z)=\frac{c(1+z)}{H_0 \sqrt{-\Omega_k}}\sin{\left( 
\sqrt{-\Omega_k}\int_0^z{\mathrm{d}z'\frac{H_0}{H(z')}}\right)},
\end{equation}
where $H(z)$ is given by the Friedmann equation, 
\be\label{H}
H(z)^2= H_0^2\left\{\Omega_{m} (1+z)^3+\Omega_{
k}(1+z)^2
+\Omega_{DE}\exp{\displaystyle\left[3\int_0^z
\frac{1+w(z')}{1+z'}\mathrm{d}z'\right]}\right\},
\ee
 and $\Omega_{DE}=1-\Omega_m-\Omega_k$. Writing $D(z)=(H_0/c)(1+z)^{-1}d_L(z)$, we have, using $h(z)=H(z)/H_0$~\cite{CCB}
\ba
w(z)
&=&-\frac13\frac{\Omega_k (1+z)^2+2(1+z)hh'-3h^2}{
(1+z)^2[\Omega_m(1+z)+\Omega_k]-h^2}\label{wH}\\
&=&\frac{2(1+z)(1+\Omega_kD^2)D''-\left[(1+z)^2\Omega_kD'^2+2(1+z)\Omega_kDD'-3(1+\Omega_kD^2)\right]D'}
{3\left\{(1+z)^2\left[\Omega_k+(1+z)\Omega_m\right]D'^2-(1+\Omega_kD^2)\right\}D'}\, ,
\label{w}
\ea
which, in principle, gives the dark energy EOS from Hubble rate or distance data provided we know $\Omega_m$ and $\Omega_k$. Written in this way we see just how difficult characterising deviations from $\Lambda$ is: we need second-derivatives of distance data. If we wish to reconstruct $w(z)$ in a meaningful way, we must know distances extremely accurately~-- it is likely in fact that there are large classes of models we can never rule out using observations of the background model alone. In Fig.~\ref{fig:w} we show a caricature of this problem wherein as we go from $D(z)\to H(z)\to w(z)$ fluctuations grow by a factor of 10 and then 100. (Although the curves here look amusingly deranged, in, e.g., LTB void models, the radial profile can oscillate with no restrictions a priori. None of these can be ruled out just because they look funny! Because the distances at large $z$ match up, the CMB should be reasonably unaffected; however, there might be a large ISW effect depending on how the sound speed is treated.)
The reason for this of course is that at each step a derivative is taken; roughly speaking this induces a fluctuation of amplitude $\mathcal{O}(\Delta z^{-1})$ where $\Delta z$ is the width of the change in $D(z)$, relative to LCDM. In terms of errors, if we know $D(z)$ on a scale $\Delta z$, the error on each derivative picks up a factor of at least $\Delta z^{-1}$. 
To further complicate matters, the reconstructed $w(z)$ relies on knowing the parameters $\Omega_m$ and $\Omega_k$ accurately. For one of the dark energy models in Fig.~\ref{fig:w} we show how errors in $\Omega_{m,k}$ propagate into errors on $w(z)$: even given perfect $D(z)$, this uncertainty translates into almost catastrophic uncertainty in $w(z)$ for $z\gtrsim1$.

Depending on the class of models one is interested in, the situation for determining $w(z)$ isn't necessarily so dire as this suggests, because we can also observe perturbations, through the Bardeen potentials $\Phi$ and $\Psi$. These are related to $w(z)$, but in a way which depends on the model in question: a quintessence or k-essence model, a modified gravity model, or a void model, all with the same effective background $w(z)$ (from distance observations, say) will have a different perturbation spectrum. This would usually result from the difference in how the sound speed is treated. However, if we are reconstructing the dark energy from first principles, the sound speed also needs reconstructed. Crudely, an effective sound speed may be found from perturbations of Einstein's Field Equations as
\be
c_s^2=\frac{\delta p^\text{eff}}{\delta\rho^\text{eff}}=-\frac{\Psi''+2\mathcal{H}\Psi'+\mathcal{H}\Phi'+\left(2\mathcal{H}'+\mathcal{H}^2-\frac{1}{3}k^2\right)\Phi+\left(\frac{1}{3}k-K\right)\Psi-4\pi Ga^2w\Gamma}{3\mathcal{H}\Psi'+(k^2-3K)\Psi+3\mathcal{H}^2\Phi}\,
\ee 
where $\Phi$ and $\Psi$ are the Bardeen potentials ($\Phi$ is the `time' one, and $\Psi$ is the space one here), $\mathcal{H}$ is the conformal Hubble rate and $'$ is derivative wrt conformal time; $K$ is the curvature, and $k$ is the wavenumber. The gauge invariant $\Gamma$ is zero for adiabatic perturbations, and proportional to the entropy flux for non-adiabatic modes; for alternative descriptions of dark energy which involve an effective fluid, this would have to be determined. To determine the sound speed from this, we would need to know second derivatives of the growth function~-- a daunting task, but not impossible with future surveys. Alternatively, one could attempt to calculate it from background observations using
\be
c_s^2=\frac{\dot p^\text{eff}}{\dot\rho^\text{eff}}=\frac{1}{3}\frac{(1+z)\left(hh''+h'^2+\Omega_k\right)-2hh'}{hh'-\Omega_k(1+z)},
\ee 
which would give a/the total effective sound speed, relative to the background dynamics. Of course, in a general dark energy model, a background adiabatic sound speed ($c_s^2=\dot p^\text{eff}/\dot\rho^\text{eff}$) may not be the same as the perturbative one ($c_s^2=\delta p^\text{eff}/\delta\rho^\text{eff}$). This is a further potential consistency test for dark energy models. It deserves further investigation to see how to do this properly in a sufficiently general way.

\subsection{Curvature?}

One of the key problems for reconstructing $w(z)$ lies in measuring the parameters $\Omega_k$ and $\Omega_m$ accurately enough. For example, let's say inflation is right, and use the prior $\Omega_k=0$. What effects would this have on $w(z)$ if the actual value we should be using is non-zero? 
\begin{figure}[htbp]
\includegraphics[width=0.8\textwidth]{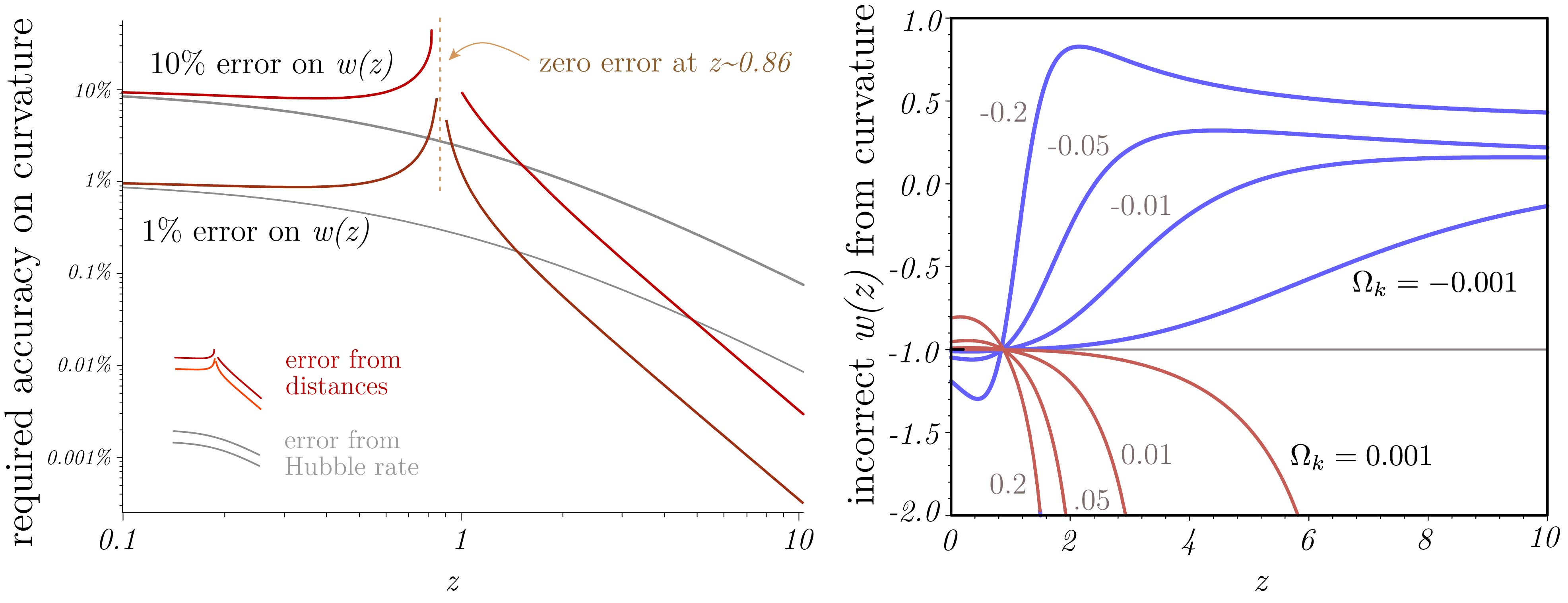}
\caption{We can mistake curvature for evolving dark energy. If the underlying model is curved LCDM and we reconstruct $w(z)$ with $\Omega_k=0$ from distances, we get the `dark energy' models, right. Conversely, if $w(z)$ really looks like one of these curves, we could mistake this for curvature by assuming $\Lambda$. The accuracy we must know the curvature to avoid problems like this is shown left; note the sweet spot at $z\sim0.86$ where we can reconstruct $w(z)$ accurately from distances irrespective of curvature. From~\cite{CCB}.}
\label{fig:curve}
\end{figure}
In Fig.~\ref{fig:curve} we show how curvature looks like evolving dark energy if we assume the wrong prior. For closed models in particular, the resulting models look rather like the effective $w(z)$ found in void models, discussed below. If we reconstruct the same curves using $H(z)$ data we find the same sort of behaviour from the dynamical effect of curvature, but the curves don't converge and cross $w=-1$ as they do for distance data, which comes from the focussing or defocussing of light by the curvature of space. At low redshift, then, both $H(z)$ and $D(z)$ suffer the same degeneracy with curvature; past the sweet spot, where dark energy may be determined from distances without contamination from curvature, they behave in opposite directions, implying that the two observables are complimentary at high redshift. Looking at it another way, if we parameterise $w(z)$ by a family of constants $w_i$ at low $z$, the degeneracy axes in parameter space will be parallel for low $z$ and will rotate as we increase $z$. See~\cite{CCB,HCCB} for more details.

\subsection{Litmus Tests for $\Lambda$}

It is clear how difficult it's going to be to reconstruct $w(z)$ in a meaningful way, if we are to do so without genuine physically motivated models. Can we instead determine of $w(z)\neq-1$ which would be enough to signify serious problems with the standard model?

\begin{figure}[htbp]
\begin{tabular}{l|r}
\includegraphics[height=0.3\textwidth]{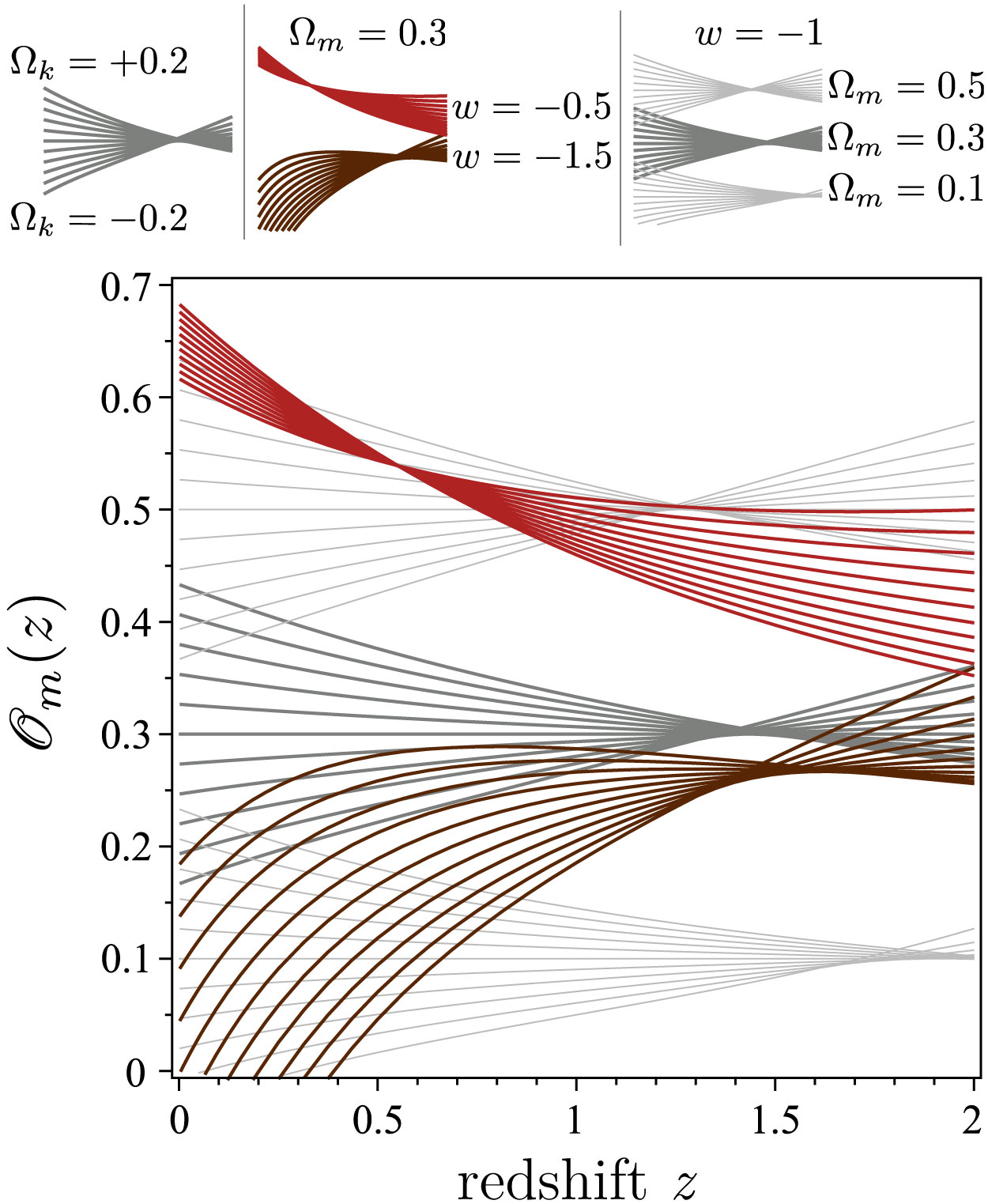} ~~~~&~~~~ 
\includegraphics[height=0.3\textwidth]{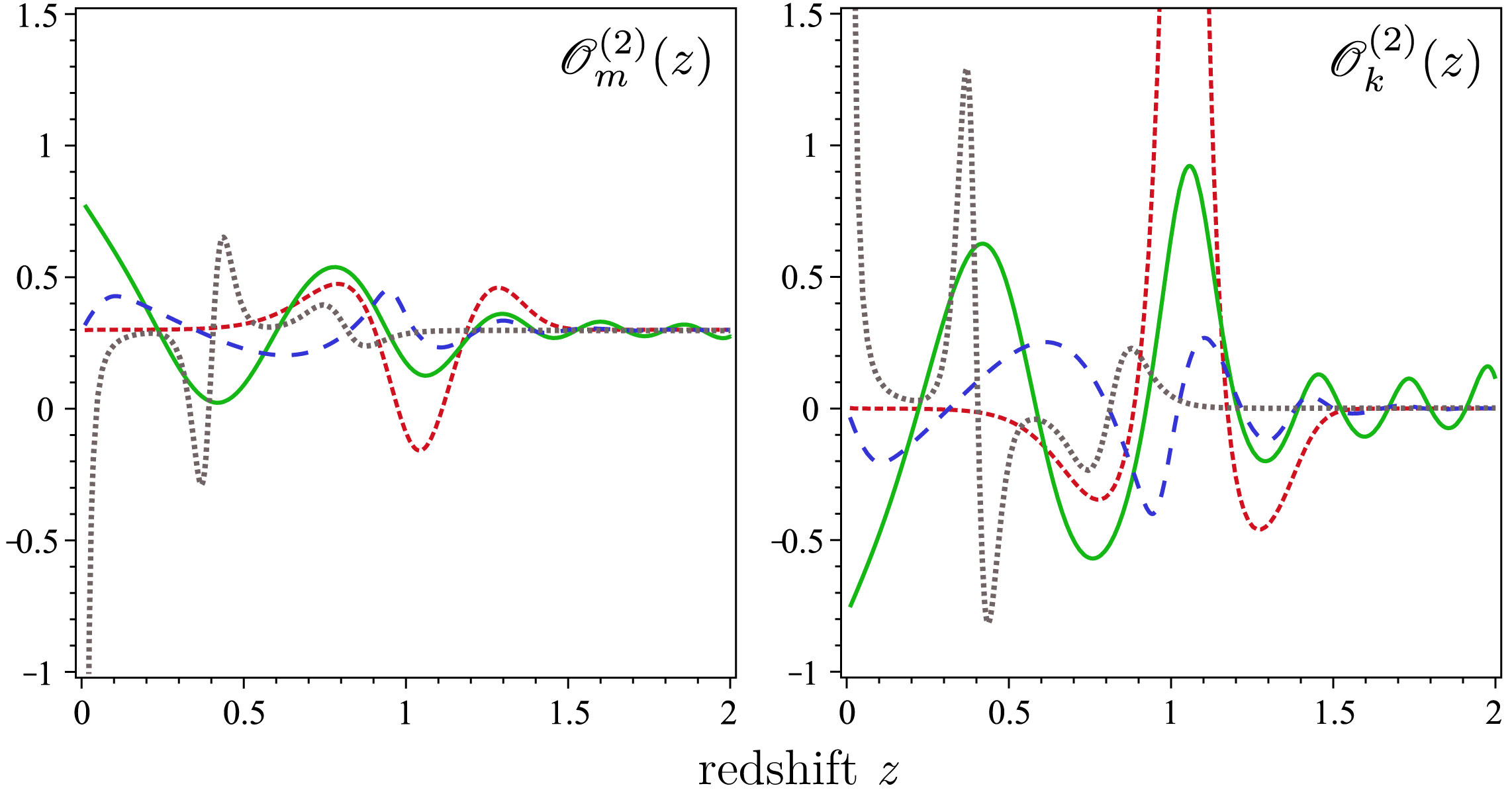}
\end{tabular}
\label{fig:oms}
\caption{Litmus tests for $\Lambda$. On the left $\mathscr{O}_m(z)$ is shown as curvature is varied, for $\Lambda$ and two constant $w$. Each fan of curves has the curvature parameter increasing from bottom to top. In the two graphs on the right we show the more general consistency tests which allow for curved models; these are shown for the `hidden' $w$'s shown in Fig.~\ref{fig:w}. Typically, $\mathscr{O}_{m,k}(z)$ are the same order as $w+1$, but may be constructed without needing any prior knowledge of $\Omega_m$ and $\Omega_k$. }
\end{figure}

Let us assume flatness for the moment. From Eqs.~(\ref{H}) and~(\ref{d_L}) we have that for LCDM~\cite{arman,ZC}
\be
\Omega_m=\frac{h^2(z)-1}{(1+z)^3-1}=\frac{1-D'(z)^2}{[(1+z)^3-1]D'(z)^2}\equiv\mathscr{O}_m(z),
\ee
where $h(z)=H(z)/H_0$ is the dimensionless Hubble rate. Viewed in terms of the observable functions on the rhs, these equations tells us how to measure $\Omega_m$ from $H(z)$ or $D'(z)$; furthermore, if flat LCDM is correct the answer should come out to be the same irrespective of the redshift of measurement. This provides an important consistency test of the standard paradigm~-- deviations from a constant of $\mathscr{O}_m(z)$ at any $z$ indicates either deviations from flatness, or from $\Lambda$, or from the FLRW models themselves~-- see the left panel of Fig.~\ref{fig:oms}. 

More generally, if we don't restrict ourselves to $\Omega_k=0$, we have that 
\ba
\Omega_{{k}}&=&\Upsilon(z)\left\{ 2\left( 1-(1+z) ^{3}
 \right)D''+3D'^{3}(D'^2-1)( 1+z)^{2}
\right\}\equiv\mathscr{O}^{(2)}_k(z)\\
\Omega_m&=&2\Upsilon(z)\left\{ \left[(1+z
 )^{2} -D^{2}-1 \right]D'' -
 \left( D'^2 -1 \right)  \left[( 1+z)D' -D 
 \right] 
\right\}\equiv\mathscr{O}^{(2)}_m(z)\label{O2k}\\
\text{where~~~}\Upsilon(z)^{-1}&=& 2\left[ 1-(1+z) ^{3} \right]  D^2 D'' 
  \nonumber\\ && -
 \left\{  (1+z)\left[  ( 1+z)^{3}-3( 1+z)+2 \right]  D'^{2}-
 2\left[ 1-( 1+z)^{3} \right] DD' -3(1+z)^{2}D^{2} \right\}D' .
\ea
The numerator of the formula for $\Omega_k$ forms the basis of the test presented in~\cite{ZC}. Again, these formulae for $\Omega_m$ and $\Omega_k$ can be used to test consistency of the LCDM model. Note that each of these tests depend only on $D(z)$, and not on any parameters of the model. Although it requires second derivatives, this makes it only as hard as measuring $w(z)$ itself, but \emph{without the degeneracy with the density parameters} which comes from estimating $w(z)$ directly using Eq.~(\ref{w}). See Fig.~\ref{fig:oms} for an illustration of these quantities.

Finally we note one last test of this sort, if we allow ourselves access to independent Hubble rate measurements as well. In curved LCDM models, the quantity 
\ba
\mathcal{L}_\mathrm{gen}(z)\!\!&=&\!\!\left\{2[(1+z)^3-1] D^2D''-[z^2(3+z)(1+z)D'^2+2[(1+z)^3-1] DD'-3(1+z)^2D^2]D'\right\}h(z)^2
\nonumber\\
\!\!&+&\!\!2[(1+z)^3-1]  D
-3(1+z)[3(1+z)D^2-z^2(z+3)]D',
\ea
is zero, which may be found by substituting Eq.~(\ref{OK}), below, into Eq.~(\ref{O2k}). If we combine distance data with Hubble rate data and find this to be non-zero, we know something is wrong with $\Lambda$.

\section{Small inhomogeneity and backreaction}

\begin{figure}[htbp]
\begin{tabular}{c}
\includegraphics[width=0.8\textwidth]{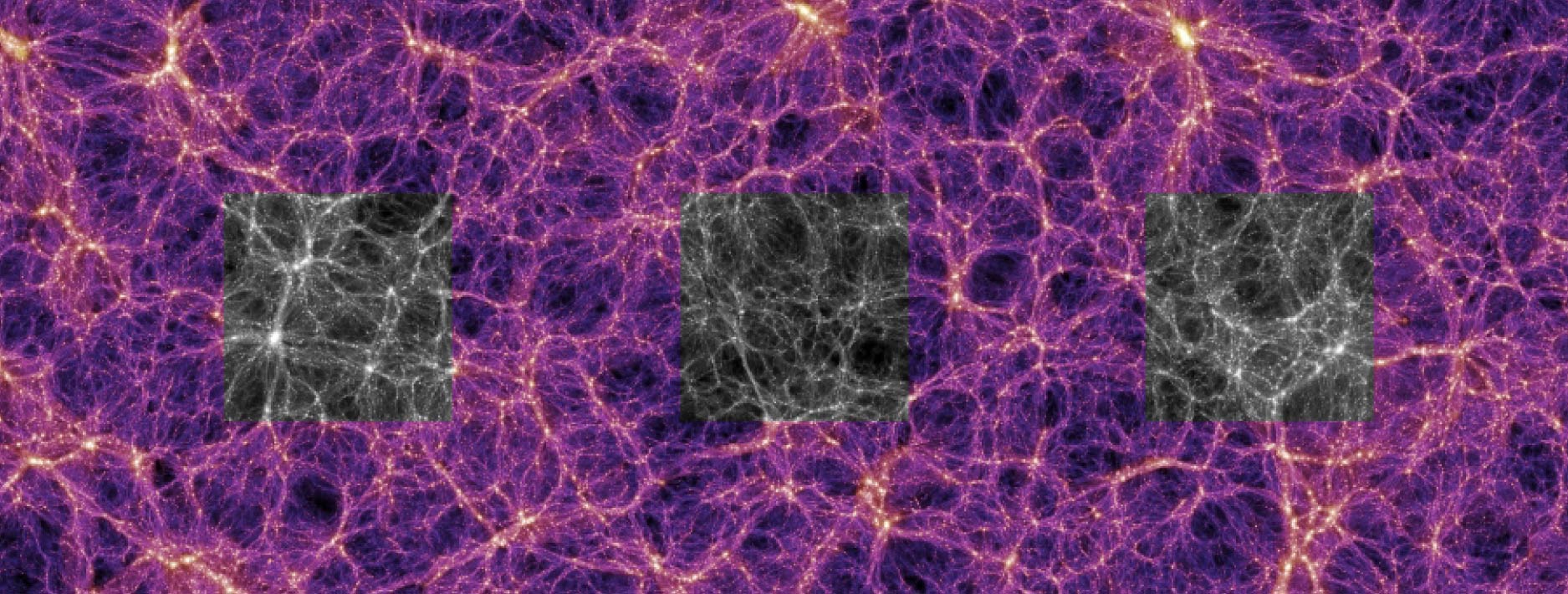}\\
\begin{tabular}{cccc}
\includegraphics[width=0.198\textwidth]{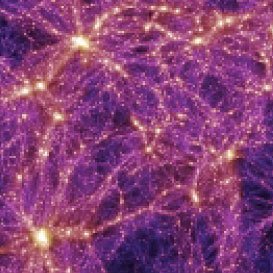}\!\!\!\!&\!\!\!\!
\includegraphics[width=0.198\textwidth]{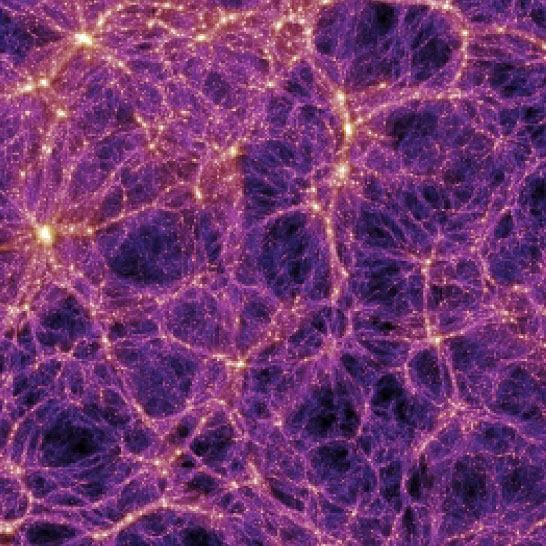}\!\!\!\!&\!\!\!\!
\includegraphics[width=0.198\textwidth]{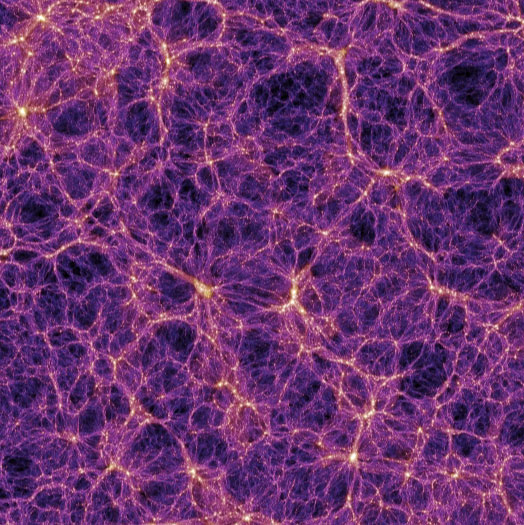}\!\!\!\!&\!\!\!\!
\includegraphics[width=0.198\textwidth]{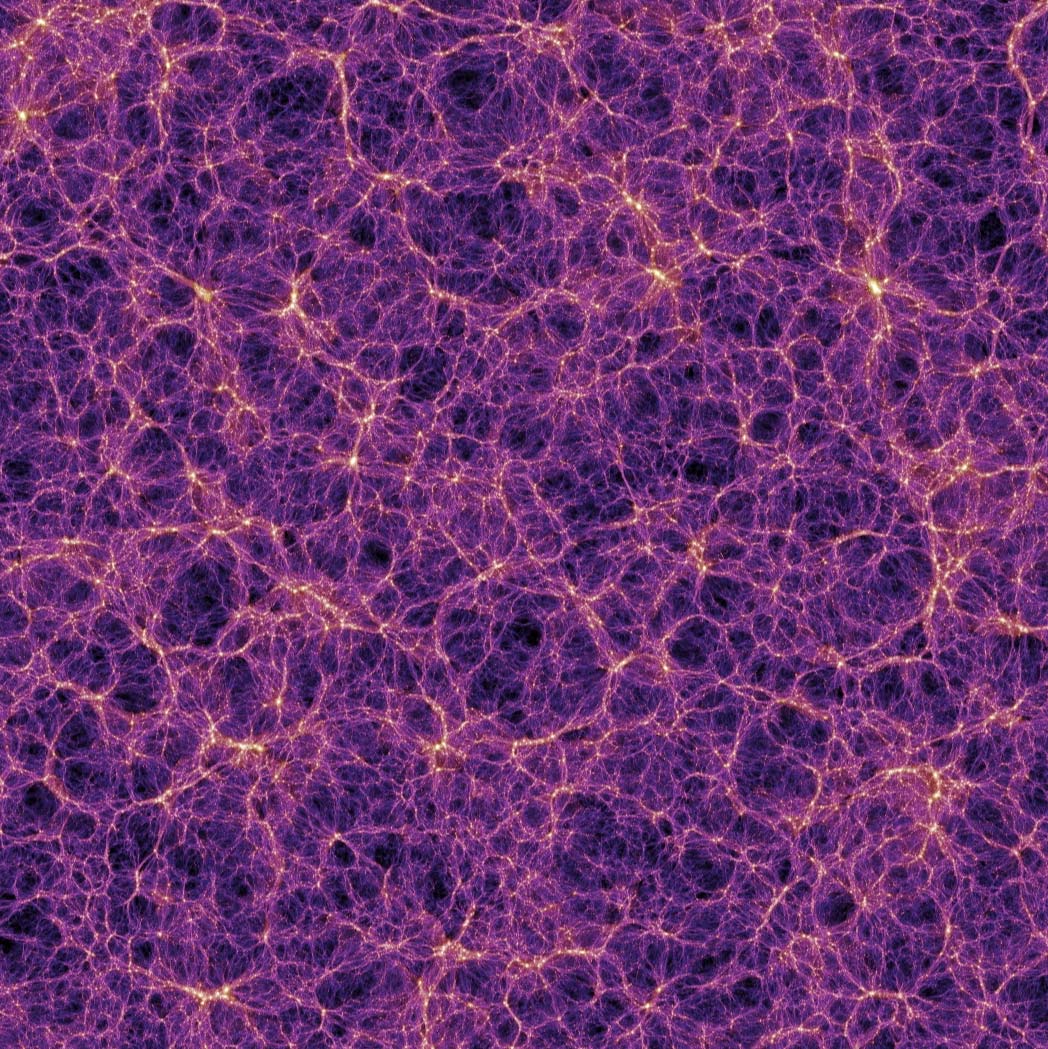}
\end{tabular}\\
\includegraphics[width=0.81\textwidth]{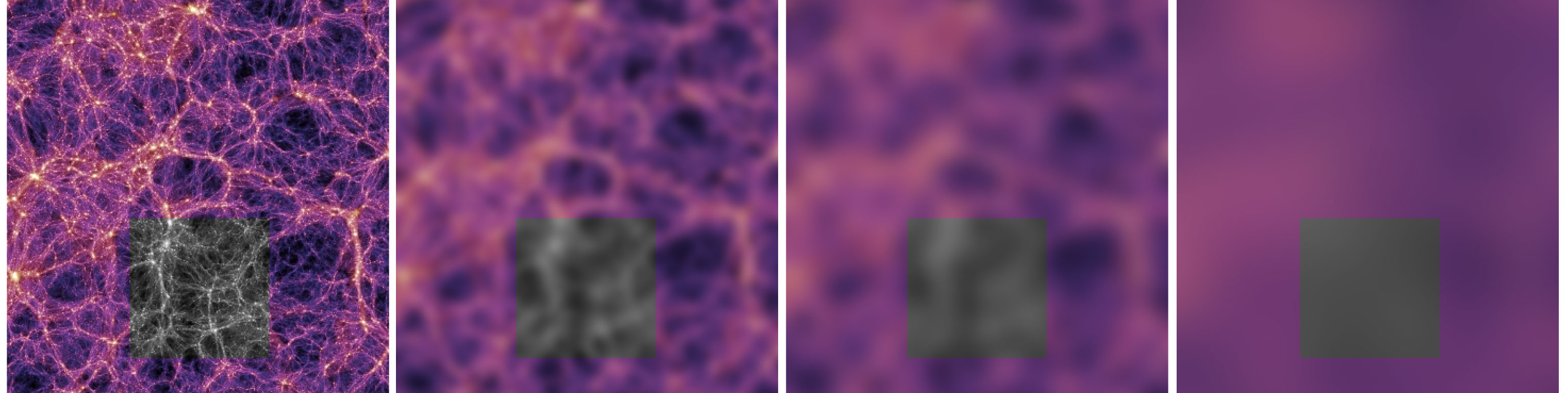}
\end{tabular}
\caption{Fiddling with the Millennium simulation~\cite{mil}. Describing the universe as `smooth' doesn't look right on scales of order $100h^{-1}$Mpc, shown here in the black and white boxes (top panel). In the central row, we zoom out from $100h^{-1}$Mpc by a factor of 2 each time (zooming out from the top left corner); only when we get to the last two boxes does it start to look homogeneous-ish, which is $\sim800h^{-1}$Mpc. (These boxes have the same depth, 15$h^{-1}$Mpc, and it's the volume that really counts, however.) The averaging problem is shown in the bottom row: how do we go from left to right? Does this process give us corrections to the `background', or is it the `background' itself?}
\label{fig:sim}
\end{figure}

A potentially critical issue arises when we try to specify and interpret a background solution at all. If we write down an FLRW solution, say, then what is the relation of that to our real universe? For example, it is natural to describe CDM (for example) as a dust fluid with zero pressure. If CDM is particulate, then the particles involved are tiny and a dust fluid should be a great description. However, most of the CDM has clustered into galaxies which have frozen out of the cosmic expansion, and the galaxies themselves act as tracers for the expansion, and so are effectively the particles of our cosmic fluid. The real metric we should use if we describe our fluid as CDM particles is fantastically complicated, and would not have a FLRW form unless smoothed in some way. Instead, it is common to think of our fundamental particles in cosmology as galaxies themselves, or clusters of them. Can we describe these accurately as as dust fluid? Perhaps. But the number of galaxies in our Hubble sphere is only of order $10^{10}$, which is \emph{well} below Avogadro's number. Looking at a simulation on 100 Mpc scales, Fig.~\ref{fig:sim}, the dark matter doesn't look like a smooth fluid at all. Does such a small number really allow the smooth  approximation  the FLRW model requires? 

Furthermore, each galaxy carries with it Weyl curvature which conveniently `averages' into Ricci curvature~-- or zero~-- when the background FLRW model is constructed. How? This is important because when we treat `a galaxy' as a particle with zero pressure, we implicitly pull into that description its gravitational field. That is, the energy density of a box of galaxies is more than the sum of the individual mass densities, simply because gravity gravitates; their own gravitational field must be somehow added to their mass-energy when considered collectively in this way. How do we do this? This is analogous to the case when we average over a gravitational wave and give it an effective energy-momentum tensor, which is then fed back into the field equations.

One aspect of this `averaging problem' comes when we try to match the late time universe today, which is full of structure, to the early time universe, which isn't. At the end of inflation we are left with a universe with curvature, $k_\text{inf}\simeq0$, and cosmological constant, $\Lambda_\text{inf}$, which are fixed for all time (and might be zero), and perturbations which are of tiny amplitude and well outside the Hubble radius; there is no averaging problem at this time, and the idea of background plus perturbations is very natural and simple to define. Fast-forward to today, where structures are non-linear, are inside the Hubble radius, and many have broken away from the cosmic expansion altogether. We may still apparently describe the universe as FLRW plus perturbations to high accuracy; that is, it is natural and seemingly correct to define a FLRW background, \emph{but it is implicitly assumed that this background is the same one that we are left with at the end of inflation}, in terms of $k_\text{inf}$ and $\Lambda_\text{inf}$. Mathematically we can follow a model from inflation to today, but when we try to fit our models to observations to describe our local universe we are implicitly smoothing over structure, and this can contaminate what we think our inflationary background FLRW model should be. Indeed, it is not clear that the background smoothed model should actually obey the field equations at all; nor is it trivial to calculate the average null cone, and how this compares with the path taken by light in an averaged model. Within the standard paradigm the averaging problem also becomes a fitting problem; are the background parameters we are fitting with the CMB actually the same as those when fitting SNIa? 

If there were no dark energy problem, this would be mainly a technical question of interest for internal consistency of our model. With the dark energy problem it becomes paramount for two reasons. One is that it could provide the `solution', in that apparent acceleration results from our inability to model non-linear structure properly. The second is because of the difficulties we have in trying to reconstruct $w(z)$~-- as we have seen, even tiny changes to $D(z)$, for example, can result in big changes to $w(z)$, so if our background model is off by just a little in an unknown way, chaos could ensue in our interpretation of things.

\subsection{Smoothing the standard model}

Take the metric in the Poisson gauge,
\be\label{metric}
\d s^2=-\left(1+2\Phi+\Phi^{(2)}\right)\d t^2+a^2\left(1-2\Phi-\Psi^{(2)}\right)\delta_{ij}\d x^i\d x^j\ ,
\ee
which describes a flat FLRW background with cosmological constant plus density perturbations up to second-order. $\Phi$ obeys the usual Bardeen equation, and $\Phi^{(2)},\Psi^{(2)}$ may be given as integrals over $\Phi^2$ terms. Ideally, we would like to construct a smooth FLRW `spacetime' from an lumpy inhomogeneous spacetime by averaging over structure. This would, in principle, have a metric 
\begin{equation}
\label{EffMetric}
\d s_{\text{eff}}^{2}=-\d\tau^{2}+a_{\CD}^{2}\gamma_{ij}\d y^i\d y^j\mbox{ ,}
\end{equation}
where $\tau$ is the cosmic time and $a_{\CD}(\tau)$ an averaged scale factor, the subscript $\CD$ indicating that it has been obtained at a certain spatial scale~$\CD$, which is large enough so that a homogeneity scale has been reached; in this case $\gamma_{ij}$ will be a metric of constant curvature.  Unfortunately, we don't know how to construct $\d s_{\text{eff}}^{2}$. We don't know what field equations it would obey, nor do we know how to calculate observational relations; none of these would be as in GR.

What we can do, however, is calculate averages of scalars associated with Eq.~\ref{metric}, such as the Hubble rate, and deceleration parameter. There are more complications lurking, however. When we calculate the averaged Hubble rate, we have to decide what to calculate the Hubble rate \emph{of}, and on which spatial surfaces to perform the average. We could calculate the average of the fluid expansion in the rest-frame of the fluid. This seems like a natural choice except that, in the case of perturbed FLRW, the gravitational field has a different `rest-frame' from the fluid. This may be characterised by the frame in which the magnetic part of the Weyl tensor vanishes and the electric part becomes a pure potential field, with potential $\Phi+\Phi^2+\frac{1}{4}(\Phi^{(2)}+\Psi^{(2)})$. This is the frame $n_a=-N\partial_at$, where $N^2=-g_{tt}=\left(1+2\Phi+\Phi^{(2)}\right)$. The fluid with 4-velocity $u^a$ drifts through this frame with peculiar velocity $v^a=(0,v^i)/a$ where  $v_{i}=\frac{1}{2}\partial_{i}(2v^{(1)}+v^{(2)})$. An observer at rest with respect to the gravitational field will measure the fluid to have expansion
\be
\theta=(g^{ab}+n^an^b)\nabla_a u_b\ .
\ee 
In the lumpy spacetime, when we consider the length-scale $\ell$ associated with $\theta$, we have 
\[
\frac{1}{3}\theta=n^a\nabla_a\ln\ell=\frac{1}{\ell}\frac{\d \ell}{\d t_\text{prop}}=\frac{1}{N\ell}\frac{\partial \ell}{\partial t}.
\]
We have a freedom in our choice of coordinates in the lumpy spacetime to be those which are most appropriate in the smoothed one.
Hence, if we demand that $t$ represents the proper time in the smoothed spacetime, $\tau$, we can define a pre-synchronised, smoothed, Hubble parameter using $N\theta$ as~\cite{juls,CAL}
\begin{equation}
\label{Eq:Hubble}
H_{\CD}\equiv \frac{1}{3}\average{N\theta}=\frac{1}{3V_{\CD}}\int_{\CD}J\d^{3}x\, N\theta \mbox{ ,}
\end{equation}
where $J=a^{3}\left[1-3\Phi+\frac{3}{2}\left(\Phi^{2}-\Psi^{(2)}\right)\right]$ is the 3-dimensional volume element and $V_\CD=\int_{\CD}J\d^{3}x$ is the volume of the domain.
We may think of this as the average Hubble parameter which preserves the length-scale $\ell$ after smoothing, according to the pre-chosen proper time in the smoothed spacetime. We can use this to then define the effective scale factor for the averaged model as the function $a_{\CD}(t)$ obeying:
\begin{equation}
\label{Eq:SFdef}
H_{\CD}=\frac{\partial_{t}a_{\CD}}{a_{\CD}}\mbox{ .}
\end{equation}
We then define an averaged deceleration parameter as
\be\label{q}
q_\mathcal{D}(z)=-\frac{1}{H_\mathcal{D}^2}\frac{\ddot{a}_\mathcal{D}}{a_\mathcal{D}},
\ee
where ${\ddot{a}_\mathcal{D}}/{a_\mathcal{D}}$ is given by a generalised Raychaudhuri equation.

To calculate $H_\CD$ perturbatively for the metric~(\ref{metric}) we may expand the averages $\average{\cdot}$. For any scalar function $\Upsilon$, the Riemannian average $\langle \Upsilon\rangle_{\CD}$ can be expanded in terms of the Euclidean average over the domain $\CD$, defined as 
$ \langle \Upsilon \rangle =\frac{\int_\CD\d^3x\, \Upsilon }{\int_\CD \d^{3}x}$,
  on the background space slices as:
\begin{equation}
\label{ExpandAverage}
\average{\Upsilon}=\Upsilon^{(0)}+\langle\Upsilon^{(1)}\rangle+\langle\Upsilon^{(2)}\rangle+3\left[\langle\Upsilon^{(1)}\rangle\langle\Phi\rangle-\langle\Upsilon^{(1)}\Phi\rangle\right]\mbox{ ,}
\end{equation}
where $\Upsilon^{(0)}$, $\Upsilon^{(1)}$ and $\Upsilon^{(2)}$ denote respectively the background, first order and second order parts of the scalar function $\Upsilon=\Upsilon^{(0)}+\Upsilon^{(1)}+\Upsilon^{(2)}$.

With these definitions in mind we can now calculate average quantities in the perturbed concordance model. The averaged Hubble rate as defined by equation (\ref{Eq:Hubble}) is given by~\cite{CAL}: 
\begin{eqnarray}\label{avH}
H_{\mathcal{D}} &=& H -\<\dot{\Phi} \> -\frac{2(1+z)^2}{9  H^2 \Omega_{m}}\left(H\<\partial^2{\Phi} \>+\<\partial^2\dot{\Phi} \>\right)
+\<\Phi~\dot{\Phi}\>\nonumber\\&&
 +\frac{2(1+z)^2}{9 H^3 \Omega_{m}^{2}}\left\{2H\Omega_{m}\left[H\<\Phi~\partial^2\Phi\> +\<\Phi~\partial^2\dot\Phi\>\right]\right.\nonumber\\
&& \left.
 +(1+3\Omega_m)H^2\<\partial^{k}{\Phi}~\partial_{k}{\Phi}\>+(2+3\Omega_m)H\<\partial^{k}{\Phi}~\partial_{k}\dot{\Phi}\>
+\<\partial^{k}\dot{\Phi}~\partial_{k}\dot{\Phi}\>
 \right\}
\nonumber\\&&
-3\<{\Phi} \>\<\dot{\Phi} \>
-\frac{2(1+z)^2}{3 H^2 \Omega_{m}}\left[H\<{\Phi} \>\<\partial^2{\Phi} \>+\<{\Phi} \>\<\partial^2\dot{\Phi} \>\right]
-\frac{1}{2}\< \dot{\Psi}^{(2)} \> +\frac{1}{6}(1+z)\< \partial^2 \upsilon^{(2)} \>.
\end{eqnarray}
Here, $H=H(z)$ is the usual background Hubble rate, in terms of the background redshift. As we can see the averaged Hubble rate is a bit ludicrous. A similar expression for the Raychaudhuri equation takes up most of a page! 

Given a particular realisation of a universe we can now calculate $H_\CD$ for a given domain size $\CD$ at a particular location. This is a bit of a pain, and doesn't really tell us that much. Instead, if our perturbations are Gaussian we can calculate the ensemble averages of the averaged quantities. This would tell us, assuming ergodicity, what a typical patch $\CD$ would look like. We can also calculate the variance of a given quantity straightforwardly. 

Most of the terms we are dealing with are scalars schematically of the form $\partial^m\Phi(\bm x)\partial^n\Phi(\bm x)$ where $m$ and $n$ represent the number of derivatives (not indices), such that $m+n$ is even. Then the ensemble average relates all terms involving products of $\Phi$ to the primordial power spectrum. For example:
\be\label{av-phi}
\overline{\<\partial^m\Phi(\bm x)\partial^n\Phi(\bm x)\>}
=(-1)^{(m+3n)/2}\int \d k\, k^{m+n-1}\mathcal{P}_\Phi(k).
\ee
For integrals which result from $m+n>4$ we have a UV divergence which needs a cutoff; this necessitates a smoothing scale $R_\mathcal{S}$. Other terms of the form $\overline{\<\cdots\>\<\cdots\>}$ have an explicit dependence on the length scale $R_\CD$, specifying the radius of our spherical domain.

\begin{figure}[tbp]
\includegraphics[width=0.8\textwidth]{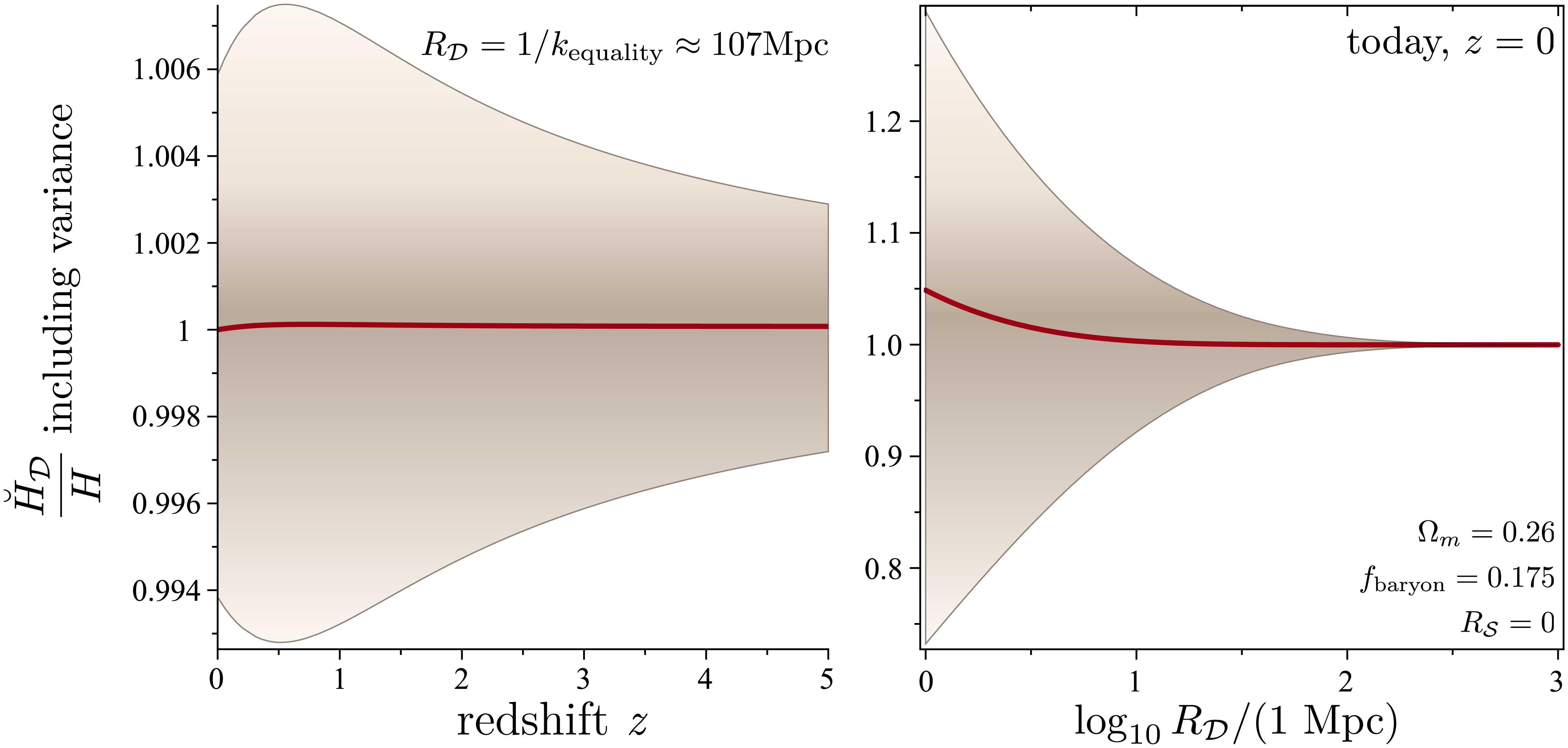}
\caption{The change to the Hubble rate from backreaction, from second-order scalar perturbations. On the left, we see the backreaction effect growing over time until $\Lambda$ kicks in. The right shows the change today as we change the domain size; only when the domain is around the equality scale does the backreaction effect become managable. }
\label{fig:H}
\end{figure}
In Fig.~\ref{fig:H} we show the averaged Hubble rate for the concordance model. As a function of the background redshift parameter the backreaction grows during the matter era, and starts to decay as $\Lambda$ becomes important. The variance is significantly larger than the pure backreaction, implying that in a domain of size $k_\text{equality}^{-1}$ we can expect fluctuations in the Hubble rate of order a percent or so. On domains smaller than this the backreaction and variance can become very significant indeed.

\begin{figure}[tbp]
\includegraphics[width=1.0\textwidth]{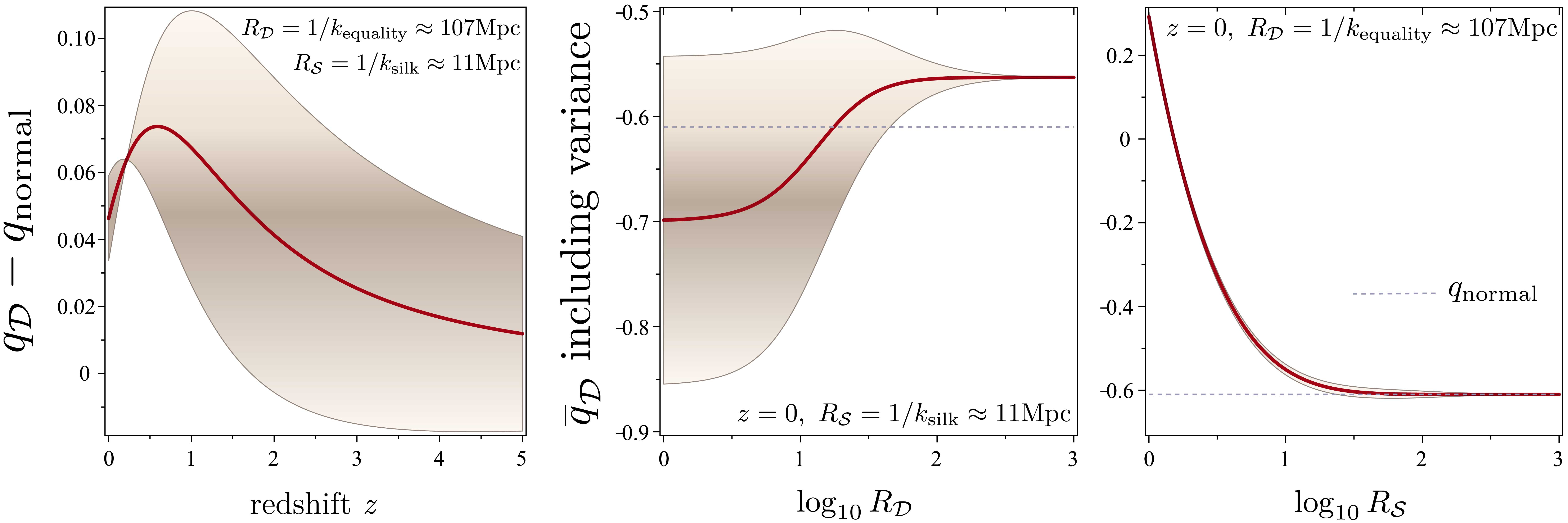}
\caption{The change to the deceleration parameter from backreaction, from second-order scalar perturbations. On the left, we see the backreaction effect growing over time until $\Lambda$ kicks in, as we saw for $H$. The variance become small for small $z$, with a sweet spot at $z\approx0.2$. The middle figure shows the change today as we change the domain size; for large domains the backreaction effect leaves a significant positive offset to the deceleration parameter. On the right we show the UV divergence which we get from the smoothing scale. }
\label{fig:q}
\end{figure}
Turning now to the deceleration parameter we see that the backreaction itself is very significant: $q$ can be increased by 10\% or more on domains of order the equality scale. In particular, the smoothing scale determines the overall amplitude, owing to the UV divergence in the integrals over the power spectrum. A much more accurate treatment of the high-$k$ part of $\mathcal{P}_\Phi$ is warranted to get a more accurate estimate; however, choosing the Silk scale seems to be a very conservative choice for the smoothing scale, and this still implies a very important effect. 

In the middle graph of Fig.~\ref{fig:q}, we see that as we increase the domain size the variance drops to zero, and the overall backreaction effect does not vanish, even on Hubble scales. This is important because our background model is renormalised by the process of perturbing then smoothing. It is worth comparing these results to Fig.~\ref{fig:sim}, the middle row in particular. To calculate the deceleration parameter accurately we must consider domains of $\sim 600$Mpc across (i.e., the third box, roughly), and then somehow subtract off the structure appropriately. This is critical for dark energy reconstruction because we don't know how to subtract the backreaction parts of our model to get at the underlying background we are left with at the end of inflation. How will this affect our reconstructed $w(z)$? We don't yet know. 

In this volume, Jim Peebles argues that the backreaction effect must be small, and $\sim\<\partial_k\Phi\partial^k\Phi\>$~\cite{peebles}. This seems to be correct as far as the Hubble rate goes. But when we calculate the deceleration parameter from the generalised Raychaudhuri equation, this has terms like $a\<\partial^2\Phi\partial^2\Phi\>\propto\langle\delta^2\rangle$ which dominate; the ensemble average of this has a UV divergence for a scale-invariant spectrum, and conservatively cutting this off at the Silk scale gives a significant contribution. The origin of this term it to be found in the divergence of the velocity between the rest frames of the gravitational potential and the CDM.

\section{Large inhomogeneity: voids and the Copernican Principle}

An odd  explanation for the dark energy problem in cosmology is one where  the underlying geometry of the universe is significantly inhomogeneous on Hubble scales, and not homogeneous as the standard model assumes. These models are possible because we have direct access only to data on our nullcone and so can't disentangle temporal evolution in the scale factor from radial variations. Such explanations are considered ungainly compared with standard cosmology because naively  they revoke the Copernican principle, placing us at or very near the centre of the universe. Perhaps this is just because the models used~-- Lema\^itre-Tolman-Bondi (LTB) or Szekeres to date~-- are very simplistic descriptions of inhomogeneity, and more elaborate inhomogeneous ones will be able to satisfy some version of the Copernican principle (CP) yet satisfy observational constraints on isotropy (e.g., a Swiss-Cheese model or something like that). 

Alternatively, within the multiverse context, one can imagine a vast universe in which our little patch happens to be rather inhomogeneous. 
We can even imagine a multiverse in which there are many void-like regions; even if we happened to be near the centre of one with a Hubble-scale inhomogeneity, this may be natural within a larger context, in the same way we discovered that the Milky Way is not particularly special once understood in the context of a plethora of galaxies. With this idea, we needn't violate the Copernican Principle if we live near the centre of a Hubble scale void; rather, we should just change our perspective~\cite{uzan}. Indeed, as argued in~\cite{uzan}, it is worth reflecting on the fact that the anthropic `explanation' for the current value of $\Lambda$, which relies on a multiverse of some sort for its philosophical underpinning, necessitates the violation of the Copernican principle simply because the vast majority of universe patches are nothing like ours, and not at all suitable for complex life. 

Instead, we may think of these models as smoothing all observables over the sky, thereby compressing all inhomogeneities into one or two radial degrees of freedom centred about us~-- and so we needn't think then as ourselves `at the centre of the universe' in the standard way. In this sense they are a natural first step in developing a fully inhomogeneous description of the universe.  

Whatever the interpretation, such models are at the toy stage, and have not been developed to any sophistication beyond understanding the background dynamics, and observational relations; in particular, perturbation theory and structure formation is more-or-less unexplored, though this is changing. They should, however, be taken seriously because we don't yet have an  explanation for dark energy in which the late time physics is well understood in any other form. Indeed, one could argue that these models are in fact the most conservative explanation for dark energy, as no new physics needs to be introduced.

We model an inhomogeneous void as a spherically symmetric LTB model with metric
\ba
\label{LTBmetric2}
ds^2 = -dt^2 + \frac{\apar^2(t,r)}{1-\kapr}dr^2 + \aperp^2(t,r)r^2d\Omega^2\,,
\ea
where the radial ($\apar$) and angular ($\aperp$) scale factors are related by 
$\apar \equiv (\aperp r)^{\prime}$
and a prime denotes partial derivative with respect to coordinate distance $r$. The curvature $\kappa=\kappa(r)$ is not constant but is instead a free function.
From these two scale factors we define two Hubble rates:
\ba\label{H}
\Hperp= \Hperp(t,r) \equiv \frac{\dot a_\perp}{\aperp},~~~~\Hpar=\Hpar(t,r) \equiv \frac{\dot a_{\|}}{\apar}
\ea
using which, the Friedmann equation takes on its familiar form:
\ba
\frac{\Hperp^2}{\Hperpo^2}=\omegmo\aperp^{-3} + \omegko \aperp^{-2},
\ea
where $\omegmo(r)+ \omegko(r)=1$ and $\omegmo(r)$ is a free function, specifying the matter density parameter today. In general, $\Hperpo(r)$ is also free, but removing the decaying mode fixes this in terms of $\omegmo(r)$. 

The fact that these models have one free function implies that we can design models such that they can give any distance modulus we like. For example, if we choose $\omegmo(r)$ to reproduce a LCDM $D(z)$ then the LTB model is a void with steep radial profile which is, strictly speaking, non-differentiable at the origin if we want $q_0<0$. Much has been made of this non-differentiability, but it's irrelevant for this sort of cosmological modelling (we don't expect any model to hold smoothly down to infinitesimal scales!). However, such freedom implies that it's impossible to tell the difference between an evolving dark energy FLRW model and a void model, using distance data alone. 

\begin{figure}[htbp]
\begin{tabular}{c}
\begin{tabular}{ccc}
\includegraphics[height=0.25\textwidth]{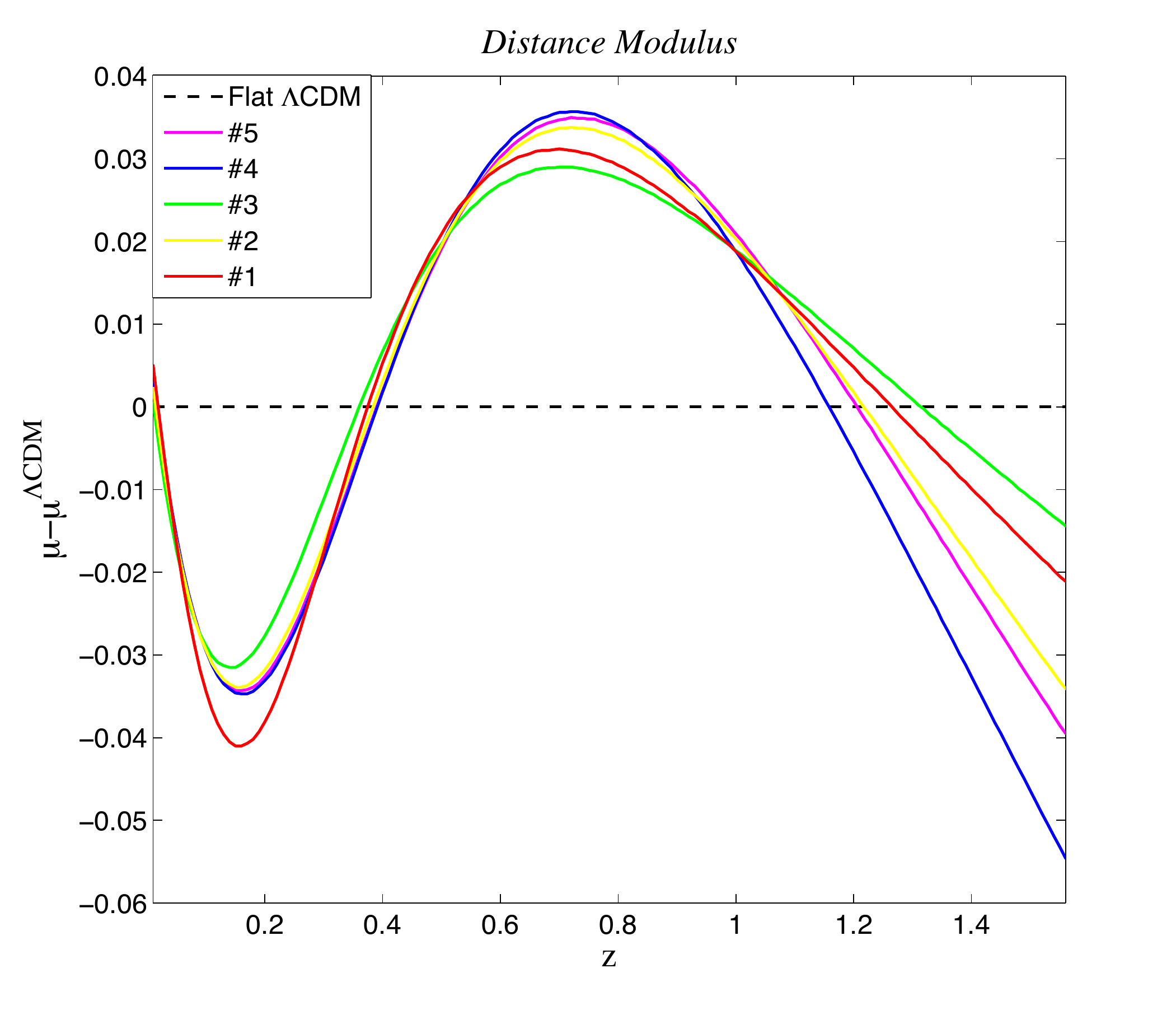}&
\includegraphics[height=0.25\textwidth]{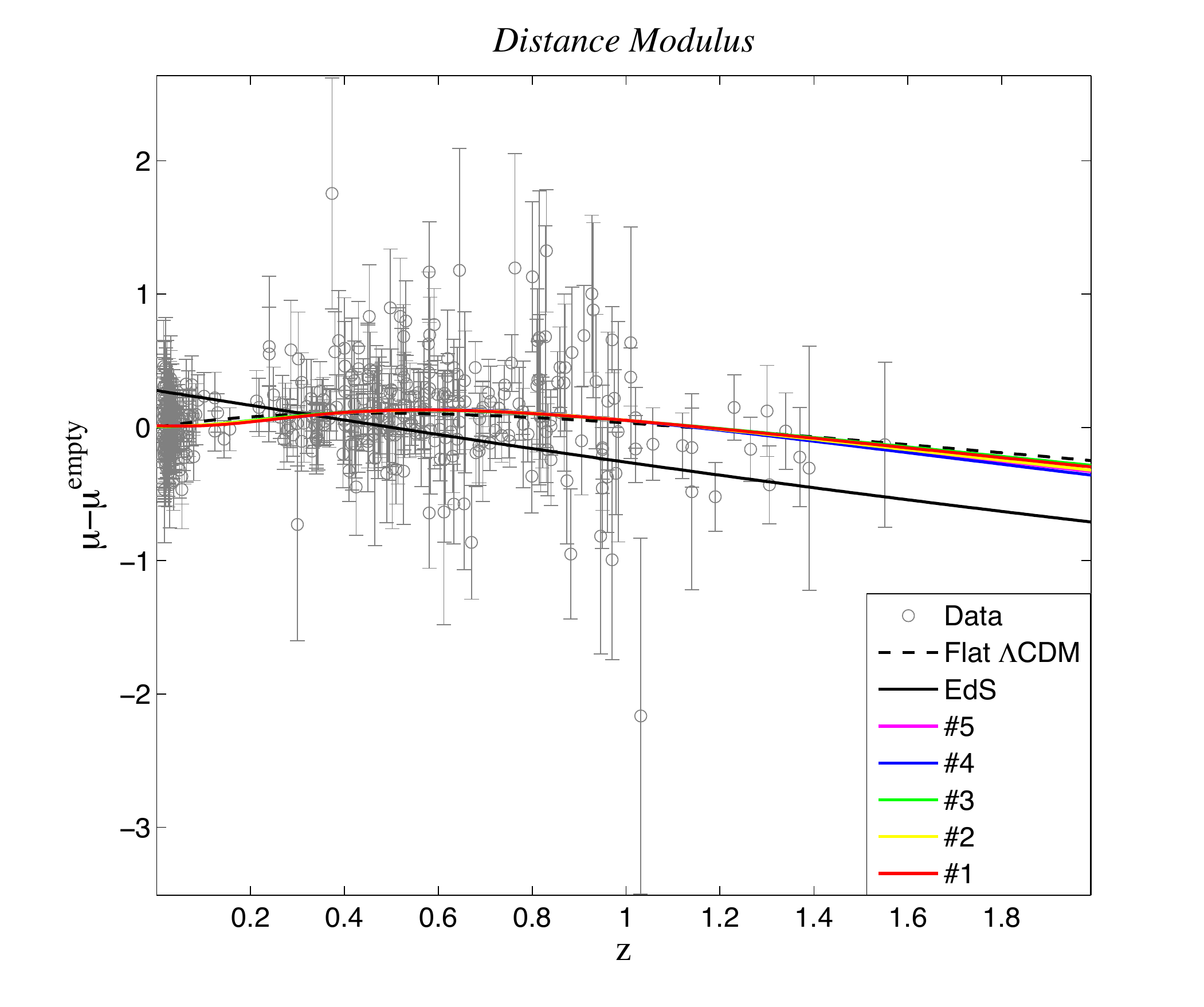}&
\includegraphics[height=0.25\textwidth]{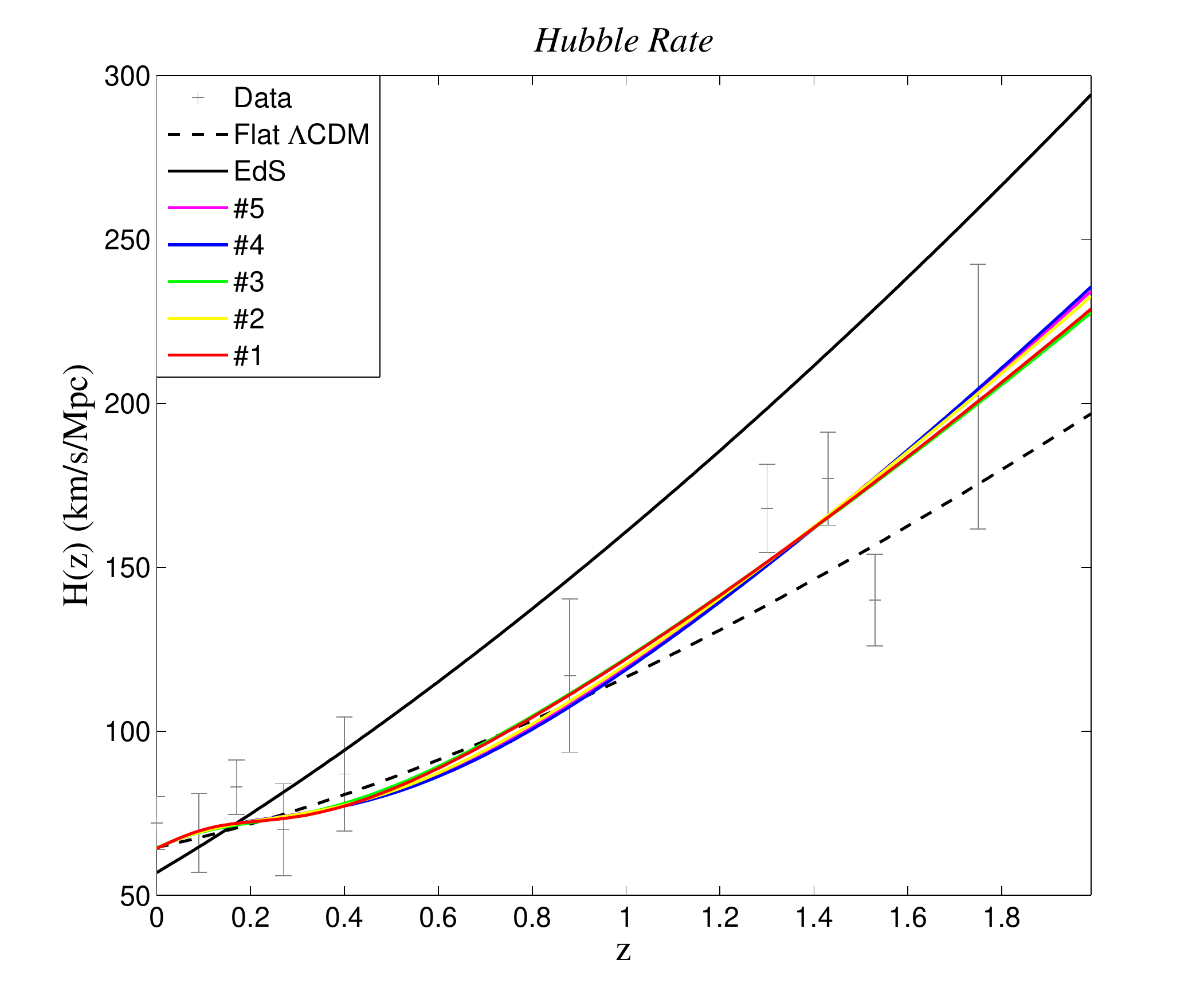}
\end{tabular}
\\
\begin{tabular}{ccc}
\includegraphics[height=0.29\textwidth]{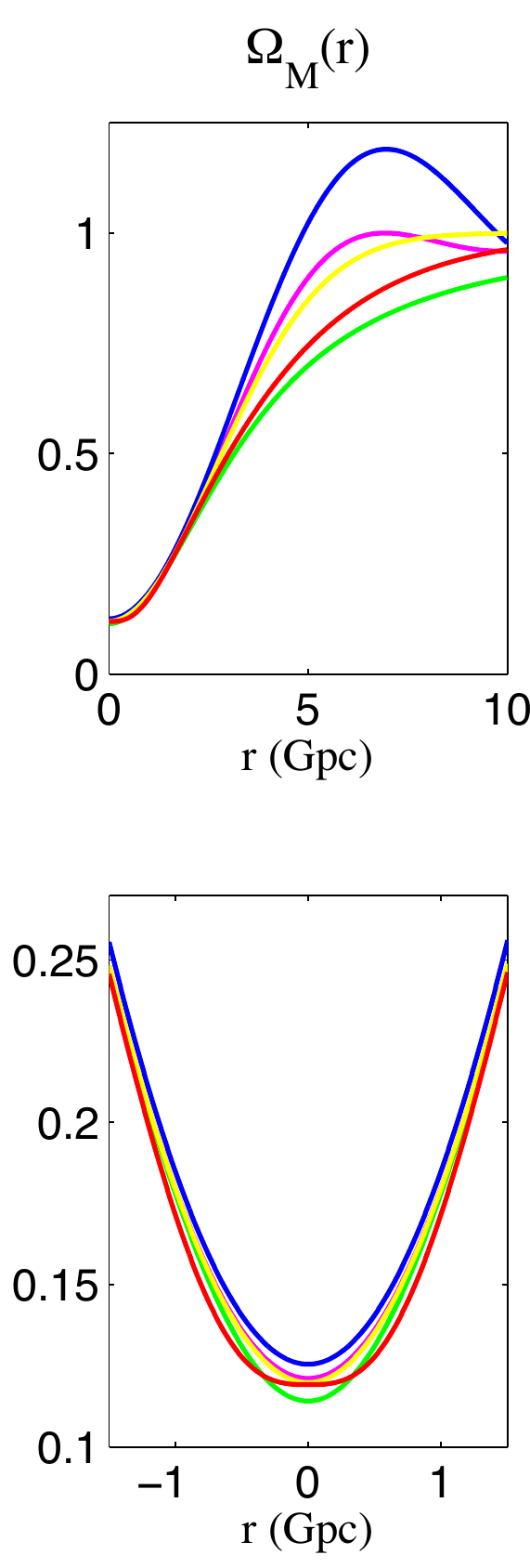}&
\includegraphics[height=0.29\textwidth]{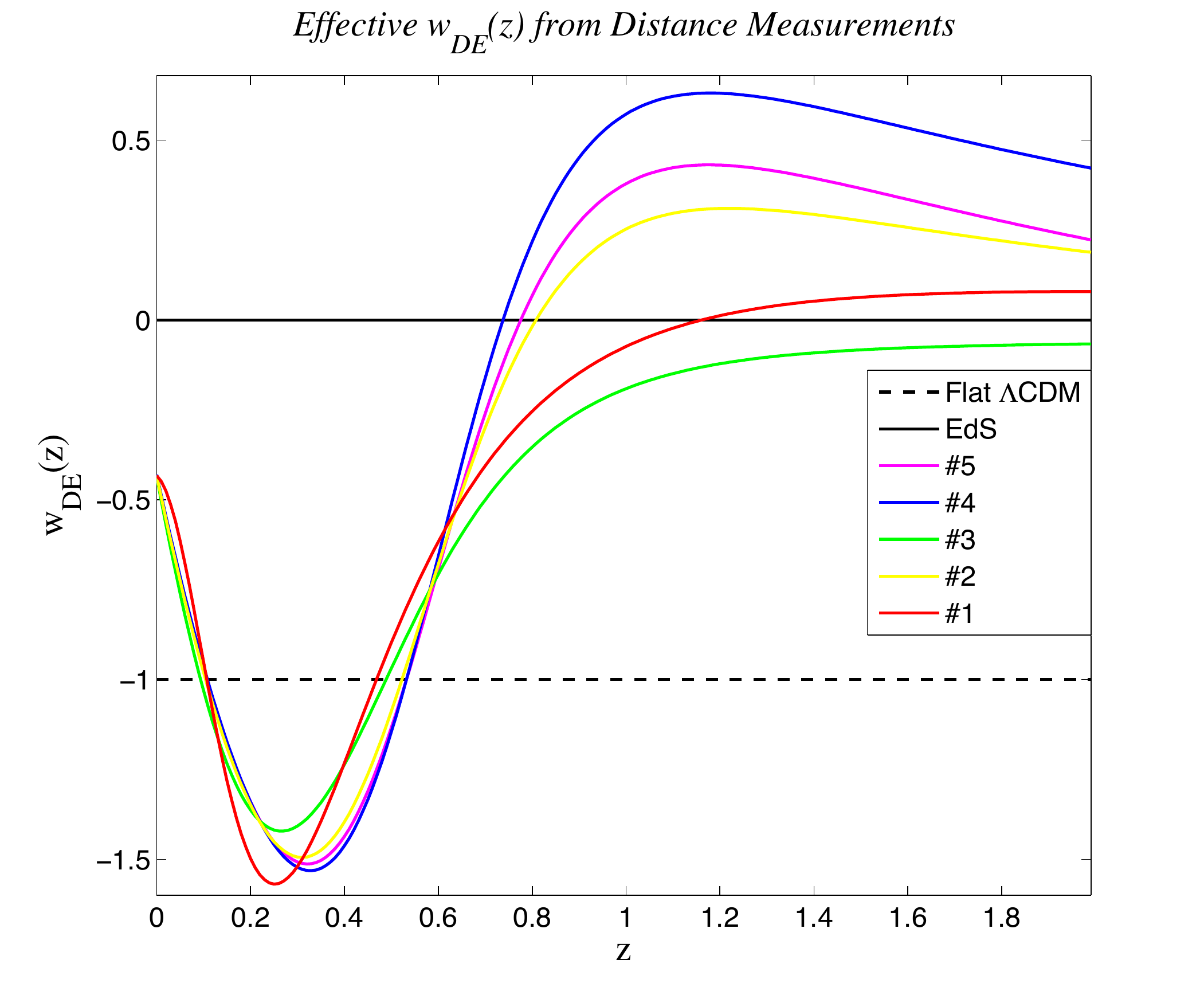}&
\includegraphics[height=0.29\textwidth]{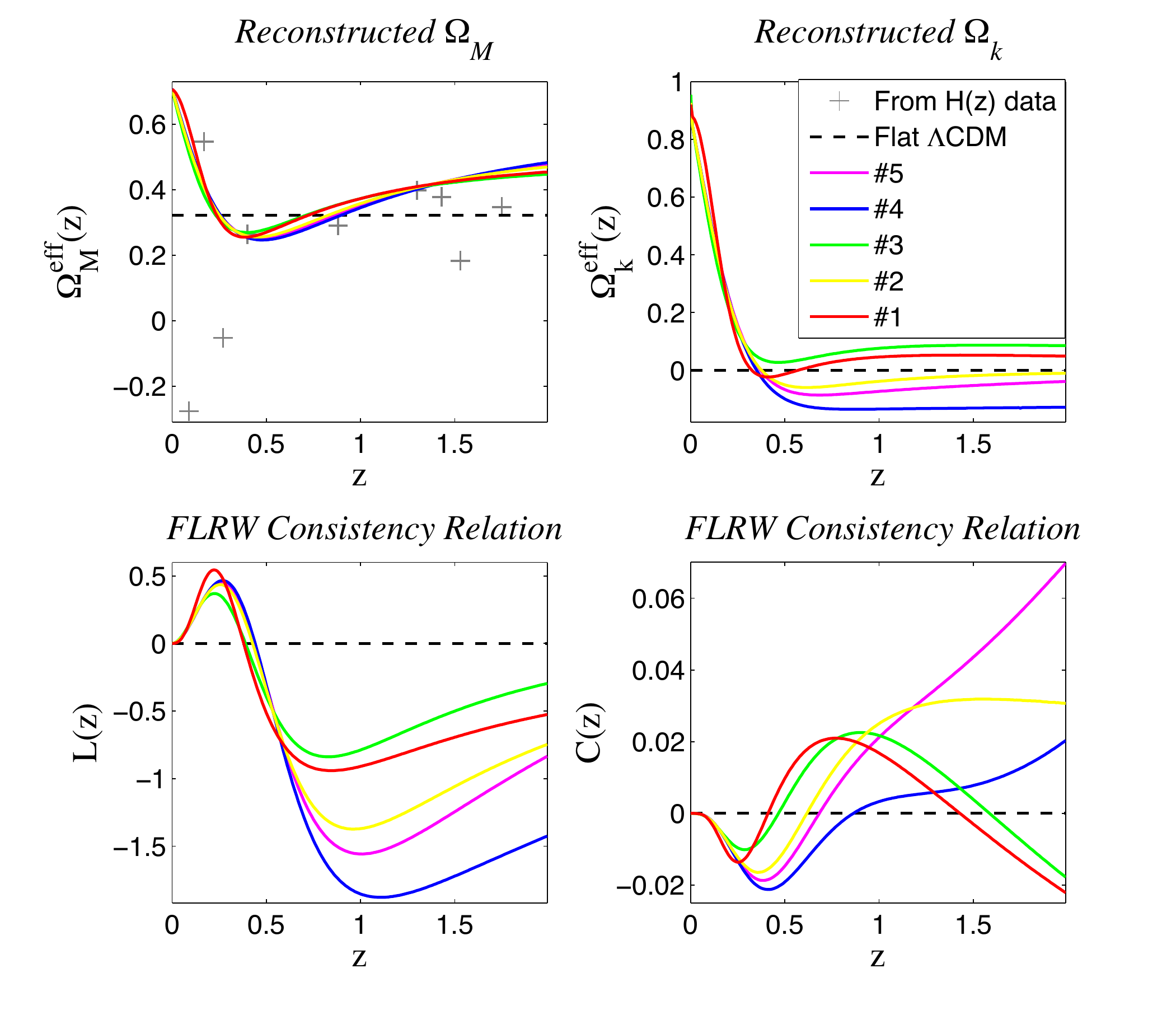}
\end{tabular}
\end{tabular}
\caption{A void with density parameter given in the plots on the bottom left, produces a distance modulus which is very close to LCDM, and fits the constitution SNIa data better; we may also fit it to age data to give $H_\|(z)$ (top right). The data does not favour a sharp void at the centre. If we try to represent the void as an effective $w(z)$ in FLRW models we get the curves on the bottom middle. Compare to Fig.~\ref{fig:w}~-- each of those effective $w(z)$'s would translate into a void profile of one sort or another. On the bottom right, we have plots of the various FLRW consistency relations for these models~-- deviations from the dotted lines show deviations from LCDM for the left two, and deviations from FLRW for the two on the right. From~\cite{sean}.}
\label{fig:void}
\end{figure}
In Fig.~\ref{fig:void} we show some void profiles which fit the constitution supernovae very nicely~-- marginally better than LCDM in fact.  While the data is not yet good enough to say too much about the shape of the void profile, we can say that the size is 6~Gpc across, roughly. We also show the effective dark energy equation of state which produces the same distance modulus as these void models but assuming an FLRW model; note that it's apparently phantom over a reasonable redshift range.

Whether one likes these models or not, we are forced to look at them further. How can we tell the difference between dark energy in an FLRW model, and inhomogeneity in a void model if they produce similar $D(z)$? One can argue that it must be `unlikely' that we are at the centre, and so look for anisotropies in, for example, the CMB. This will probably constrain us to be within 1\% of the centre, say, compared to the Hubble scale. Does this rule them out because this is unlikely? Unfortunately not~--  the Copernican principle would be violated in our Hubble patch, but this would just leave us with a spatial version of the coincidence problem to explain. What we need are ways to rule out these models for a central observer. 

One way to do this is (presumably) through the matter power spectrum and the CMB; that is, by observing perturbations. Essentially, this will give generalised Bardeen potentials, but which are now all mixed up with gravitational waves and vector modes. For example, in LTB models the generalised Bardeen equation is, for even parity modes with spherical harmonic index $\ell\geq2$~\cite{CCF}:
\be
\ddot\varphi+4H_\perp\dot\varphi-2\frac{\kappa}{a_\perp^2}\varphi=S(\chi,\varsigma). 
\ee
Here $S(\chi,\varsigma)$ is a source term which couples this potential to gravitational waves, $\chi$, and vector modes, $\varsigma$~-- these in turn are sourced by $\varphi$. This represents the fact that the gravitational field is inherently dynamic even at the linear level, and that structure may grow more slowly due to the dissipation of potential energy into gravitational radiation. 
Mathematically, we have a very complicated set of coupled pdes to solve for each harmonic $\ell$. Furthermore, since $H_\perp=H_\perp(t,r)$, $a_\perp=a_\perp(t,r)$ and $\kappa=\kappa(r)$, perturbations in each shell about the centre will grow at different rates, and it's because of this that the perturbations generate gravitational waves and vector modes. These equations have not yet been solved in full generality. Thus, we can expect different structure growth in LTB models, but in what way we don't yet know. Presumably, we will need to observe the full power spectrum evolving from high redshift until today to definitely decide between FLRW and LTB, because there also exists a degeneracy with the primordial power spectrum.

Instead of resorting to perturbations, we can try to test the validity of the FLRW assumption directly using properties of the background solutions. For example, rearranging Eq.~\ref{d_L}, we have the curvature parameter today given by~\cite{CCB}
\begin{equation}\label{OK}
\Omega_k=\frac{\left[h(z)D'(z)\right]^2-1}{[D(z)]^2}\equiv\mathcal{O}_k(z).
\end{equation}
On the face of it, this gives a way to measure the curvature parameter today by combining distance data with Hubble rate data, irrespective of the redshift of measurement. In FLRW this will be constant as a function of $z$, independently of the dark energy model, or theory of gravity. Alternatively, we may re-write this as the condition that~\cite{CBL}
\begin{equation}
\mathcal{C}(z)=1+h^2\left(DD''-D'^2\right)+hh'DD'\ ,
\end{equation}
must be zero in any FLRW model at all redshifts, by virtue of Eq.~(\ref{OK}). In more general spacetimes this will not be the case. In particular, in LTB models, even for a  for a central observer, we have $\mathcal{C}(z)\neq0$, or $\mathcal{O}_k(z)\neq$const. 

This tells us that in all FLRW models there exists a precise relationship between the Hubble rate and distance measurements as we look down our past null cone. This relationship can be tested experimentally \emph{without specifying a model at all}, if we reconstruct the functions $H(z)$ and $D(z)$ in a model independent way and independently of each other. This would then provide a model-independent method by which to experimentally verify the Copernican assumption, and so verify the basis of the FLRW models themselves. 

Considering $\mathcal{O}_k(z)$ and $\mathcal{C}(z)$ in non-FLRW models reveals how useful these tests might be. In this volume, David Wiltshire discusses $\mathcal{C}(z)$ in the timescape cosmology, arguing that it may be used more broadly than just a test of the Copernican principle (see also~\cite{wilt}).  Although $\mathcal{C}(z)\neq0$ in virtually all inhomogeneous models, once isotropy about us is established on a given scale, we can use these tests to eliminate the possibility of radial inhomogeneity, thereby testing the Copernican principle.   

It should be pointed out that there exists a class of LTB models for which we can design both $D(z)$ and $H_\|(z)$ to match those of FLRW, and so these would pass the homogeneity test presented here, if it were used with $H=H_\|$. Such models utilise the decaying mode of LTB models to do this~\cite{charles}. (A similar solution with $H_\perp=H_{\Lambda\text{CDM}}$ surely exists too.) To really rule this special class out we would need to use $H_\perp(z)$ as an observable in $\mathcal{O}_k(z)$ or $\mathcal{C}(z)$ which can be measured using the time drift of cosmological redshifts, given by, in LTB~\cite{UCE}:
\be
\dot z(z)=(1+z)H_0-H_\perp(z)\ .
\ee
To really rule out LTB models, then,  we need to show that $H_\|(z)=H_\perp(z)$ for all $z$. $H_\|(z)$ can be measured using the relative ages of passively evolving galaxies, using
$\displaystyle\frac{\d t}{\d z}=\frac{1}{(1+z)H_\|(z)}\ .$
This is nice for this test because it can be used relatively model-independently. Other methods for measuring $H(z)$ include the BAO and the perturbative dipole in the distance modulus~\cite{ruth}, both of which rely on FLRW perturbations at present.

\section{Discussion}

If the cosmological constant, $\Lambda$, is the underlying cause of dark energy then the viewpoint that cosmology is nearly complete is probably true. 
If $\Lambda$ is wrong, however, then all bets are off, as `dark energy' could then be all sorts of things, from $x$-essence to modified gravity to large-scale inhomogeneity.

Attempts to justify the value of $\Lambda$ using landscape arguments along with the multiverse, necessarily combined with the anthropic principle, open two tenable doors: one is that the universe as a whole is colossal or even infinite, and the other is that our Hubble volume is both tiny and incredibly special. If this is indeed the case, then the multiverse breaks with the Copernican principle in a spectacular way: we exist in a highly exceptional corner of the universe, and are not typical observers except in our little patch where the fundamental constants are just so (which might be rather large in terms of Hubble volumes of course, but small in terms of all that there is).  If we're happy to break with a `global Copernican principle' to explain the value of $\Lambda$, it is not philosophically unreasonable to break with a `local Copernican principle' instead, in order to~-- perhaps~-- preserve a global one. The void models offer the possibility of describing dark energy as radial inhomogeneity in a dynamically predictable model, rather than as an unknown dynamical degree of freedom in a postulated homogeneous model. One can even argue that inflation predicts such a scenario~\cite{linde}! That they break with the Copernican principle on our Hubble scale is a cause for concern, and even a sophisticated inhomogeneous model which drops the spherical symmetry may well suffer a similar problem. On the scale of the multiverse anything goes, so it's easy to imagine a universe with inhomogeneous fluctuations on Hubble and super-Hubble scales~\cite{uzan}. Of course almost all speculation beyond a few times our Hubble sphere is really `fiction science' at the moment, requiring wild assumptions and extrapolations in all sorts of ways. Nevertheless, such speculation is important, and tests, such as those presented here, which can decide the scale of homogeneity out to the Hubble scale without assuming \emph{a priori} the local Copernican principle, may play an useful role in discussing such questions.

All the different possibilities for dark energy typically involve \emph{functional} degrees of freedom. From an observational point of view, we are really in the position where we have to try to rule out the very simplest models, as we can't investigate function spaces observationally in a meaningful way.
It is useful in my view to try to construct model-independent tests of different models where we can. Here I've presented a few which utilise background observations. These can probe different things: $\mathcal{O}_m(z)$ is non-zero if flat LCDM is wrong; $\mathcal{O}^{(2)}_{m,k}(z),\mathcal{L}_\text{gen}(z)$ are non-constant if $\Lambda$ is wrong, independently of all parameters of the model; and $\mathcal{O}_k(z)$ is non-constant and $\mathcal{C}(z)$ is non-zero if the FLRW models themselves are incorrect, independently of the theory of gravity used, or fluid used to model the dark energy. These are useful precisely because they can be implemented without specifying a background model at all, if the observables $D(z)$ and $H(z)$ are constructed directly from the data. 

A important issue lies in the backreaction of perturbations which is tied up with the averaging problem. How do we smooth structure to connect to the background model at all? We have seen that backreaction in the concordance model renormalises our background, which gives a change in the deceleration parameter of at least 10\%, apparently far higher than would be guessed by hand-waving arguments. Thus, even if dark energy is the cosmological constant, it might be difficult to see it as such until this problem is further quantified and understood.

\paragraph{Acknowledgements} I would like to thank Kishore Ananda, Bruce Bassett, Tim Clifton, Marina C\^ortes, George Ellis, Sean February, Ren\'ee Hlozek, Julien Larena, Teresa Lu, Mat Smith, Jean-Philippe Uzan and Caroline Zunckel  for collaboration and discussions which went into the work presented here. I would like to thank Sean February for the plots in Fig.~\ref{fig:void}, and Roy Maartens for comments. This work is supported by the NRF (South Africa).

\end{document}